\documentclass[aps,prl,reprint,groupedaddress,superscriptaddress]{revtex4-1}

\usepackage{graphicx}
\usepackage{dcolumn}
\usepackage{bm}
\usepackage{amsmath}
\usepackage{amssymb}
\usepackage{subfigure}
\usepackage{xcolor}

\begin{document}


\title{Deionization Shock Driven by Electroconvection in a Circular Channel}
\author{Zhibo Gu}
\thanks{contributed equally}
\author{Bingrui Xu}
\thanks{contributed equally}
\author{Peng Huo}
\affiliation{Department of Aeronautics and Astronautics, Fudan University, Shanghai,200433,China}
\author{Shmuel M. Rubinstein}
\affiliation{John A. Paulson School of Engineering and Applied Sciences, Harvard University, Cambridge, Massachusetts 02138, USA}
\author{Martin Z. Bazant}
\affiliation{Department of Chemical Engineering and Department of Mathematics, Massachusetts Institute of Technology, Cambridge, Massachusetts 02139, USA}
\author{Daosheng Deng}
\email{dsdeng@fudan.edu.cn}
\affiliation{Department of Aeronautics and Astronautics, Fudan University, Shanghai,200433,China}
\date{\today}

\begin{abstract}
In a circular channel passing over-limiting current (faster than diffusion), transient vortices of bulk electroconvection are observed in salt-depleted region within the horizontal plane. The spatiotemporal evolution of the salt concentration is directly visualized, revealing the propagation of a deionization shock wave driven by bulk electroconvection up to millimeter scales. This novel mechanism leads to quantitatively similar dynamics as for deionization shocks in charged porous media, which are driven instead by surface conduction and electro-osmotic flow at micron to nanometer scales.  The remarkable generality of deionization shocks under over-limiting current could be used to manipulate ion transport in complex geometries for desalination and water treatment.

\end{abstract}

\pacs{Valid PACS appear here}
\maketitle
Ion transport in electrochemical cells is essential for electrochemical energy storage, desalination for water treatment, and biomedical applications \cite{Probsteinbook, Newmanbook, schoch2008transport}. Designing complex geometries is one of the typical approaches to control ion transport, as illustrated by the stack of alternating cation- and anion-exchange membranes in classical electrodialysis (ED) \cite{Sonin1968}. More recently,  ion enrichment/depletion resulting from overlapping electric double layers  in micro/nanochannels \cite{pu2004ion-enrichment} has been applied to biomolecule separation \cite{kim2010nanofluidic}.  Polarizable porous electrodes or particles under applied voltages can also induce capacitive deionization, in a variety of geometries \cite{rubin2016induced-charge, porada2013review}.

In these and other applications, many intriguing phenomena are associated with the passage of over-limiting current (OLC), faster than diffusion, to an ion-selective membrane~\cite{nikonenko2014} or electrode~\cite{han2014overlimiting,han2016dendrite}. Physical (as opposed to chemical~\cite{andersen2012}) mechanisms for OLC fall into two general categories: bulk electroconvection (EC) associated with extended space charge on the membrane  \cite{rubinstein2000electroosmotically, zaltzman2007electroosmotic, rubinstein2008direct, Yossifon2008, handesalination2013,rubinstein2015} and surface charge (SC) effects, namely surface conduction and electro-osmotic flow (EOF), through charged microchannels or porous media leading to the membrane or electrode \cite{Dydek2011overlimiting,Leakymodel2013,deng2013overlimiting,KimPRL2015,han2014overlimiting,han2016dendrite,khoo2018}. The transient response to OLC can involve the shock-like propagation of a sharp drop in salt concentration \cite{ManitheoryLang2009, ManiexpLang2009, zangle2010theory}.  The propagation of SC-driven ``deionization shocks" (DS) in charged porous media \cite{mani2011deionization} has been exploited for water desalination and purification in the emerging process of ``shock electrodialysis" \cite{deng2013overlimiting, Leakymodel2013, deng2015water, lu2015scalable} and for control of metal growth in ``shock electrodeposition"~\cite{han2014overlimiting,han2016dendrite}.   Since EC-driven vortices can also sustain OLC by creating an extended salt depletion zone, it is interesting to explore whether EC alone can give rise to DS.

In this Letter, we report the observation of EC-driven DS in a circular microchannel. Vortices reminiscent of bulk EC are identified within the horizontal plane, and the spatiotemporal evolution of concentration is directly visualized. Propagation of EC has a remarkable agreement with the proposed model.

\begin{figure}[b]
\includegraphics[width=0.9\linewidth]{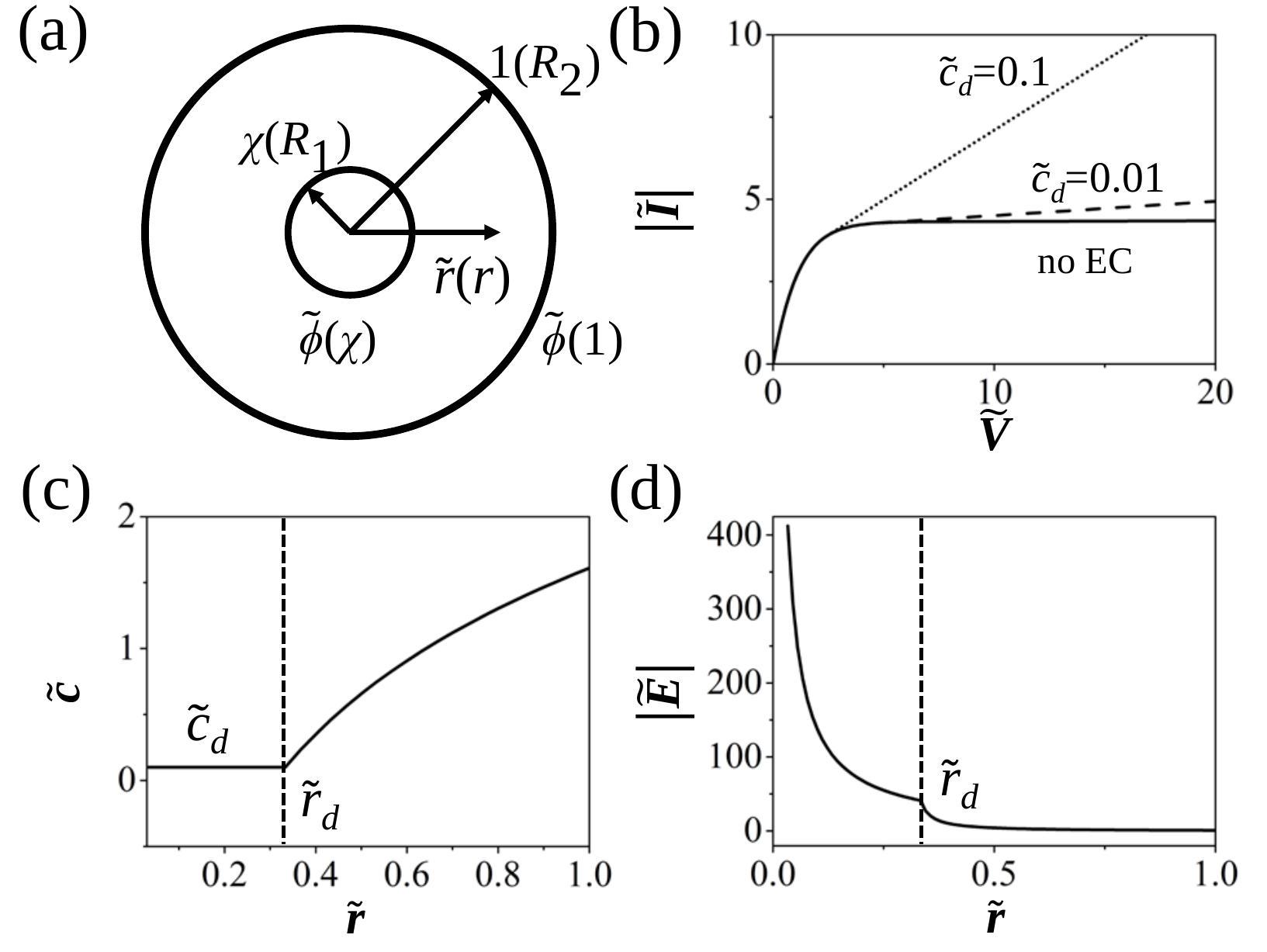}
\caption{(a) Sketch of a circular channel. The positive voltage bias for $\tilde{\phi}(1)>\tilde{\phi}(\chi), \tilde{I}<0$. (b) OLC dependent on $\tilde{c}_d$, (c) concentration distribution, and (d) profile of the electric field ($\tilde{\phi}=35, \tilde{c}_d=0.1, R_d=\tilde{r}_d R_2$).}
\label{fig:current}
\end{figure}

\begin{figure*}[t]
\includegraphics[width=0.92\linewidth]{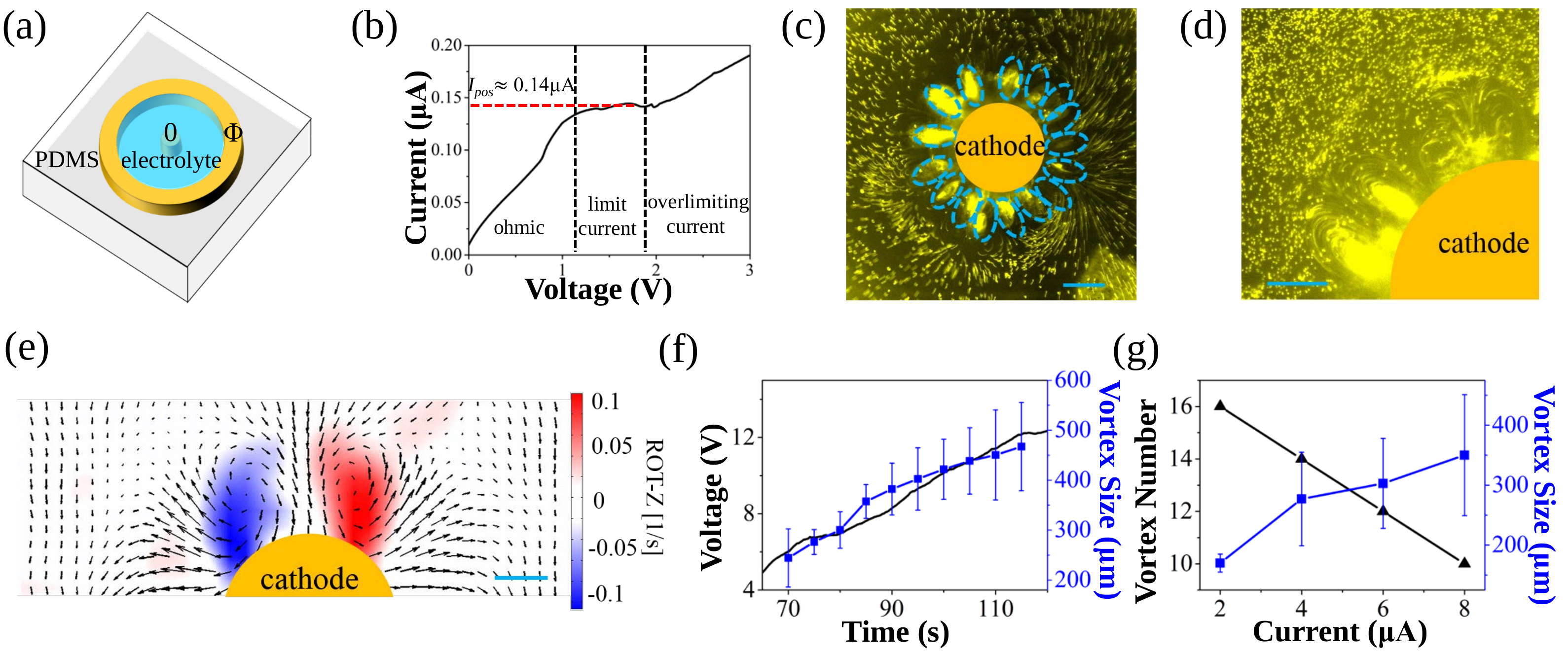}
\caption{Vortex observation from a top-down view. (a) Sketch of the PDMS device ($2R_2 = 6$ mm, $H \approx 35 \mu$m, $\chi$ = 1/30, $\Phi > 0$ for the positive voltage bias). (b) The $I-V$ curve showing $I_{pos}\approx$ 0.14 $\mu$A.  (c-e) Under an applied current at 2 $\mu$A in the overlimiting current regime, the time-lapse snapshot at t $=$ 60 s with an exposure time of 5 seconds  for (c), and with an exposure time of 100 ms for (d); (e) PIV images with a short exposure time of 40 ms  at t $=$ 20 s when the first vortex pair occurs. (f) Voltage and vortex size increase with time. (g) The number of vortex decreases with the applied current, while the size of vortex increases with the applied current when the vortex occurs. Scale bar in (c-e) for 200 $\mu$m.
\label{fig:vortex}
}
\end{figure*}

\emph{Simple model of EC-driven OLC.}---Consider the following model problem, studied experimentally below. A dilute, binary $z:z$ electrolyte with concentration ($c_0$) fills a circular channel with an inner radius ($R_1$) and outer radius ($R_2$ ) ($\chi = R_1 /R_2 < 1$) under an applied voltage (Figure \ref{fig:current}a). In the steady state, under the assumption of the azimuthal symmetry and charge neutrality, the Nernst-Planck equations are simplified into a 1D dimensionless form, equating cation flux to current density and anion flux to zero for an ideal cation-selective surface \cite{Dydek2011overlimiting}:
\begin{subequations}
	\label{Nearstplanck}
\begin{align}
	\frac{\mathrm{d}\tilde{c}}{\mathrm{d}\tilde{r}}+\tilde{c}\frac{\mathrm{d}\tilde{\phi}}{\mathrm{d}\tilde{r}}+\tilde{\sigma}_{EC}\frac{\mathrm{d}\tilde{\phi}}{\mathrm{d}\tilde{r}}=&-\frac{\tilde{I}}{2\pi \tilde{r}}, \label{modeqcation}\\
	\frac{\mathrm{d}\tilde{c}}{\mathrm{d}\tilde{r}}-\tilde{c}\frac{\mathrm{d}\tilde{\phi}}{\mathrm{d}\tilde{r}}+\tilde{\sigma}_{EC}\frac{\mathrm{d}\tilde{\phi}}{\mathrm{d}\tilde{r}}=&0, \label{modeqanion}
\end{align}
\end{subequations}
where $\tilde{c}=\tilde{c}_{+}=\tilde{c}_{-}$ is the (equal) dimensionless mean concentration of cations and anions scaled by $c_0$, $\tilde{r}$ the dimensionless radius scaled by $R_2$, $\tilde{\phi}$ the dimensionless potential scaled by the thermal voltage, $k_{B}T /ze$, and $\tilde{I}$ is the dimensionless current scaled by $zeDc_{0}$, assuming equal diffusivity $D$ for cations and anions.

In contrast to the Leaky Membrane Model ~\cite{Dydek2011overlimiting,yaroshchuk2012acis,Leakymodel2013,khoo2018}, where the residual {\it surface} conductivity sustaining OLC is provided by excess counter-ions in a charged porous medium, we have introduced a residual {\it bulk} conductivity due to electroconvection $\tilde{\sigma}_{EC}(\sigma_{EC}/c_0)$, which depends on the local electric field $E$ and becomes dominant above a critical threshold $E_c$ that is only reached in the depleted zone during OLC, e.g. via $\sigma_{EC}(E,c)=c_d(c) \tanh^{2}{(E/E_c)}$.
Similar to EOF \cite{Dydek2011overlimiting, deng2013overlimiting,KimPRL2015}, EC with the intense vortices near the dead end causes the formation of an extended depletion zone ($\tilde{r} \leq \tilde{r}_d$) with a nearly constant area-averaged concentration ($\tilde{c}_d$) \cite{rubinstein2000electroosmotically, zaltzman2007electroosmotic, rubinstein2008direct}, resulting in residual conductivity ($\tilde{\sigma}_{EC}\sim \tilde{c}_d$ for both types of ions), while convection is negligible in the bulk region far away from the vortices.

This two-region approximation can be used to solve the model, matching the concentration and potential at $\tilde{r}_d$, to obtain $I-V$ relationship [as shown in the Supplementary Materials (SM)],
\begin{equation}
	\label{IVoverlimit}
	\tilde{V}=\ln \left( 1+ \frac{\tilde{I}}{4 \pi \tilde{c}_d} \ln \tilde{r}_d\right)-\frac{\tilde{I}}{4 \pi \tilde{c}_d} \ln \left( \frac{R_d}{R_1} \right),
\end{equation}
where $\tilde{r}_d$ can be found for the given $\tilde{I}$ and $\tilde{c}_d$ by the concentration conservation. The calculated I-V curves (Figure \ref{fig:current}b) demonstrates OLC is sustained by $\tilde{\sigma}_{EC}$. Additionally, concentration profile and electric field are presented in Figure \ref{fig:current}c and d, indicating a constant $\tilde{c}_d$ and the sharp increase of electric field in the depletion region ($\tilde{r}\leq \tilde{r}_d$)(SM).

\begin{figure*}[t]
\includegraphics[width=\linewidth]{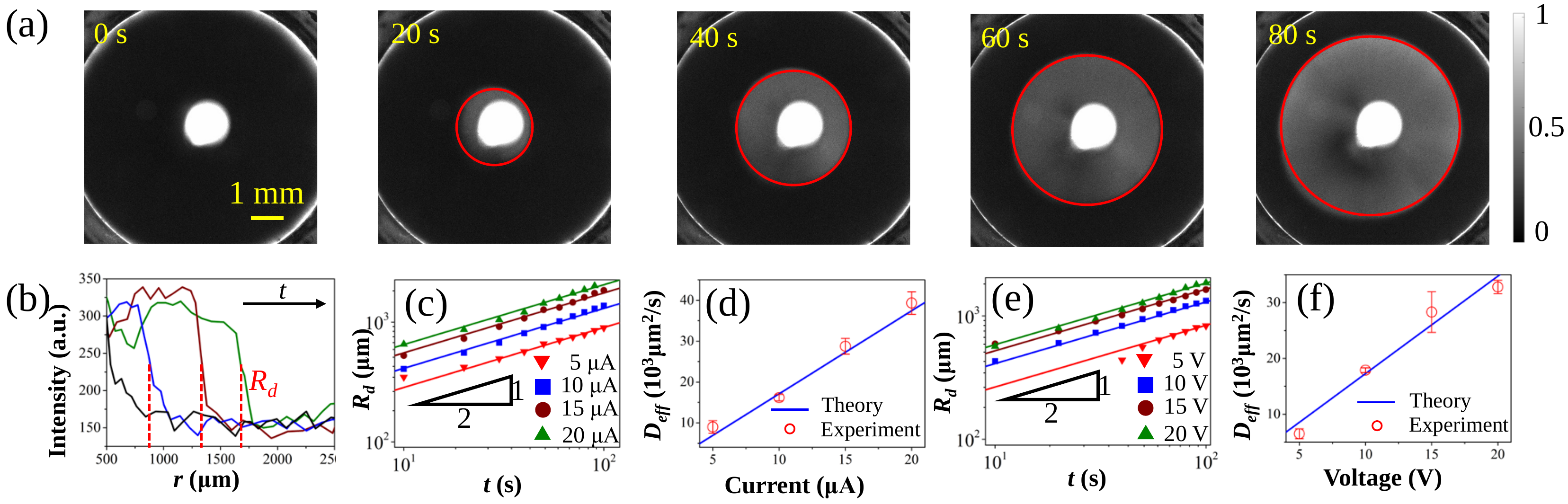}%
\caption{Spatiotemporal evolution of concentration ($R_1$ = 100 $\mu$m). (a) Snapshots of fluorescent signals at 20 $\mu$A, red circular contours for the propagation front. Saturated white color in the center due to the extra liquids around the copper wire. (b) Radial profiles of fluorescent intensity at t = 0, 20, 40, 60 sec; $R_d$ is marked by the red dash line at the steep jump of fluorescent intensity due to the concentration depletion. $R_d$  has a 1/2 power law under various currents (c) and voltages (e), while $D_{eff}$ proportional to the currents (d) and voltages (f) from Equation \eqref{frontprocurrent} and \eqref{frontprovoltage}. Error bars in (d) and (f) for the standard deviations from six measurements.}
\label{fig:concentration}
\end{figure*}

\emph{Current-voltage measurements and vortex observation.}---A PDMS device with a circular channel is illustrated in Figure \ref{fig:vortex}a, 2$R_1$ = 200 $\mu$m for the inner copper wire, 2$R_2$ = 6 mm ($\chi=1/30$) for the outer copper ring, and $ H\approx 35 \mu$m for the channel height. This combination of copper electrodes and CuSO$_4$ solution can avoid complicated chemical reactions to simplify the system \cite{deng2013overlimiting}. The measured $I-V$ curve (Figure \ref{fig:vortex}b) for 1 mM aqueous CuSO$_4$ solution under the positive voltage bias (a Keithley 2450 Source Meter) is characterized with a linear Ohmic regime, a plateau of limiting current, and OLC. Unlike the straight channel with parallel planar electrodes, $I-V$ curves and limiting currents here depend on the voltage bias (SM). By modifying PDMS surfaces to be positively charged \cite{deng2013overlimiting, methodposcharge}, the measured $I-V$ curves were uninfluenced.

As shown in Fig. \ref{fig:vortex}b, limiting current is subsequently followed by OLC, and the possible mechanism is the formation of vortex to enhance ion transport \cite{rubinstein2008direct}. Under an applied current at 2 $\mu$A the vortices are gradually generated, which are visualized by the fluorescent microscope (Zeiss, Axio Zoom V16) \cite{devalenca2015dynamics, methodflow}. At t $=$ 60 sec, the vortex is indicated by the time-lapse snapshot (Figure \ref{fig:vortex}c and d) (Video 1 in SM). By employing particle image velocimetry (PIV), the vortex fields at t $=$ 20 sec are obtained (Figure \ref{fig:vortex}e), revealing a pair of vortices exists near the cathode with velocity up to around 20 $\mu$m/sec.

We simultaneously recorded the increased voltage and built the correlation between vortex size (the vortex length) and electric response (Figure \ref{fig:vortex}f) \cite{devalenca2015dynamics}. By checking the vortices during their occurrence (Figure \ref{fig:vortex}g), vortex size increases with the current while the vortex number decreases with current, since the circumferential length is fixed for a given inner cathode.

Different from the typical bottom-up setup with the vertical concentration gradient or gravitational convection \cite{deng2013overlimiting, AliPRL2016}, here the PDMS device containing the circular channel is placed horizontally and the voltage is applied between the inner wire and outer ring (Figure \ref{fig:vortex}a), and concentration gradient is absent vertically and gravitational convection is irrelevant. In addition, unlike the typical straight microchannels with four side walls to cause EOF \cite{Dydek2011overlimiting, KimPRL2015}, here the gap of this circular channel is only 35 $\mu$m and EOF due to the bottom and top surface charges might be relevant only along the vertical direction. Again vortices were nearly unaffected by positive-charged PDMS surface, implicating the negligible role of SC mechanism and the essential role of EC.

Hence, the observed vortex confined within the horizontal plane with circumferential length up to millimeters is attributed to EC. Similar EC-driven vortices have also observed parallel-plate geometries, in terms of both size $(\sim 100 \mu$m) and velocity ($\sim 10 \mu$m/s) \cite{rubinstein2008direct, devalenca2015dynamics}, and the fastest flow is tangential to the circular cathode surface (Figure \ref{fig:vortex}c).  The vortices are thus consistent with EC instability, although it is beyond our scope  to classify the space charge as non-equilibrium    \cite{zaltzman2007electroosmotic} or equilibrium~\cite{rubinstein2015} or quantify the effect of geometrical curvature.

\emph{Concentration profile evolution.}---The spatiotemporal evolution of the EC-driven DS as visualizing through the cation concentration. The concentration of copper ions (Cu$^{2+}$) was detected by the fluorescent indicator, 20 $\mu$M Phen Green SK dipotassium salt (Invitrogen), the fluorescence intensity of which is quenched by Cu$^{2+}$ ions, \emph{i.e.}, the increased fluorescence intensity indicates the reduced concentration, and vice versa. CuSO$_4$ electrolyte at 10 mM was prepared in a compound solution (a mixture of the distilled water, thiodiethanol, and dimethyl sulfoxide) to enhance the fluorescent signals. For a constant current at 20 $\mu$A, the typical snapshots of fluorescent signals recorded by the fluorescent microscope (Zeiss, Axio Zoom V16) are shown in Figure \ref{fig:concentration}a (Video 2 in SM).

The initial homogeneous distribution of fluorescent intensity was separated into two distinct regions marked by the red circular contours (Figure \ref{fig:concentration}a), and the radial intensity is quantified in Figure \ref{fig:concentration}b, clearly demonstrating the propagation front with a sharp intensity jump. This enhanced intensity was observed only far above the limiting current, but was disappeared below the limiting current. In addition, either concentration polarization of the fluorescent indicator or electromigration of the negative-charged Phen Green SK can only cause the decreased fluorescent intensity near the inner cathode. Hence, the stronger fluorescent intensity of the inner brighter region demonstrates the significant reduction of copper concentration, reminiscent of previous observations of SC-driven DS in the same electrolyte \cite{han2014overlimiting,han2016dendrite,KimPRL2015}.  In this case, however, SC can be ruled out in favor of EC, since the concentration evolution was found to be nearly identical with a positively charged surface.

\emph{Scaling analysis.}---Propagation front ($R_{d}$) is located at the middle point of the abrupt jump, as indicated by the dash red lines (Figure \ref{fig:concentration}b). For various constant currents, $R_{d}$ is fitted by a 1/2 power-law scaling (Figure \ref{fig:concentration}c). The extracted effective diffusion coefficient ($D_{eff}$) (Figure \ref{fig:concentration}d), unlike the normal diffusion growth of depletion layer prior to the onset of EC \cite{Yossifon2008}, is about one order of magnitude higher than the typical diffusion coefficient of copper ions (0.8 $\times 10^3 \mu$ m$^2$/sec).

Physically, under the constant current (I), current in the depleted region is mainly determined by the electromigration and convection due to vortices, then ion conservation at the shock front
implies $I/(2\pi R_d) \sim \mathrm{d}R_d/\mathrm{d}t$,
\begin{equation}
R_d \sim (It)^{1/2}, \, D_{eff} \sim I \label{frontprocurrent}.
\end{equation}
Equation \eqref{frontprocurrent} shows that the square-root growth is determined by currents (far above the limiting current), and $D_{eff}$ is linearly proportional to current, consistent with the experiments (Figure \ref{fig:concentration}d).

Additionally, for various constant voltage (V), $R_d$ can be fitted by a 1/2 power-law scaling (Figure \ref{fig:concentration}e), and the extracted $D_{eff}$ increases with voltage (Figure \ref{fig:concentration}f). Ion conservation at the thin shock interface implies $\mathrm{d}R_d/\mathrm{d}t \sim V/R_d$ (electric field for the radial geometry). Then square-root growth is obtained,
\begin{equation}
R_d \sim (Vt)^{1/2}, \, D_{eff} \sim V. \label{frontprovoltage}
\end{equation}
Indeed, the experimental $D_{eff}$ is linear with the voltages (far above the onset voltage) (Figure \ref{fig:concentration}f).

Remarkably, the EC vortices do not form chaotic random patterns and remain confined to a smooth envelope during DS propagation, as shown by the red smooth circular contours in Figure \ref{fig:concentration}a.  In hindsight, a DS radially moving away from the depletion region resembles time-reversed Laplacian growth~\cite{BazantPRE2006}, which leads to smooth shock profiles~\cite{ mani2011deionization}. The stability of SC-driven DS is critical for continuous shock electrodialysis~\cite{deng2013overlimiting,lu2015scalable} and stable shock electrodeposition~\cite{han2016dendrite}, so the observation of stable EC-driven DS may lead to related applications.

\emph{Model for DS propagation.---} Although 1/2 power-law scaling of shock propagation is predicted for SC-driven DS in a circular or wedge geometry \cite{mani2011deionization}, we employ our simple physical model to understand the similar dynamics of EC-driven DS. The shock velocity is proportional to the current density \cite{mani2011deionization},
\begin{equation}
\frac{\mathrm{d}{R}_d}{\mathrm{d}t}\sim j(R_d) \sim \frac{I}{2 \pi R_d}. \label{shockmovingcur}
\end{equation}
For constant voltage ($V$), assuming the voltage approximately dropped entirely in the depletion region with a constant $c_d$ independent on time, we find (SM):
\begin{equation}
I(t) \sim \frac{c_d V}{\ln{(R_d/R_1)}}. \label{currentconstv}
\end{equation}
Then $R_d(t)$ is obtained with a fitting parameter $\alpha$,
\begin{equation}
(\frac{R_d}{R_1})^2[\mathrm{ln}(\frac{R_d}{R_1})^2-1]+1=\frac{\alpha c_dV}{R_1^2}t.  \label{rftime}
\end{equation}

\begin{figure}[t]
\includegraphics[width=0.9\linewidth]{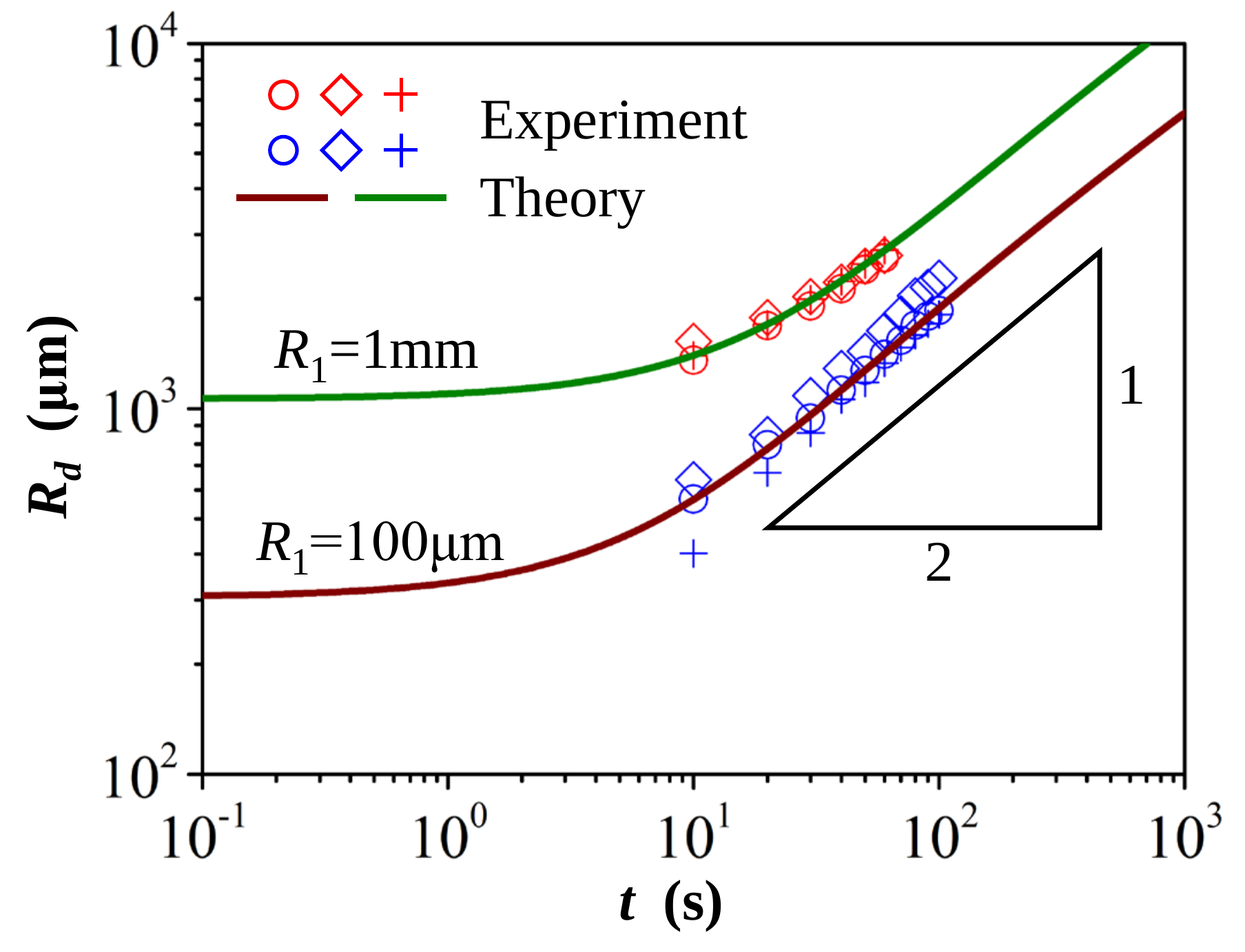}
\caption{Model for DS propagation. The experiment data under a constant voltage at 20V are fitted well by Equation \eqref{rftime}. The power law of 1/2 is recovered for a smaller radius $R_1$ = 100 $\mu$m ($\chi=1/30$), while deviation from 1/2 power law occurs for a larger radius $R_1$ = 1 mm ($\chi=1/3$). The lower bound of $R_d$ is limited by the extra liquids around the wire, and the upper bound of $R_d$ is $R_2$ (3 mm). The experimental data were reproduced by three times.} \label{fig:model}
\end{figure}

The above Equation \eqref{rftime} is in excellent agreement with the experimental data (Figure \ref{fig:model}). For a smaller radius $R_1$ = 100 $\mu$m, as shock propagates far away from the cathode ($R_d/R_1>1$), the square-root growth holds [$R_d \sim (Vt)^{1/2}$], validating the aforementioned power-law analysis. But for a larger radius $R_1$ = 1 mm, a length scale is set by $R_1$, and Equation \eqref{rftime} including the logarithms term is more accurate than the power law. Despite the simplicity, the proposed model might have captured the main features of EC-driven DS.

\emph{Discussion.---}
Theoretically, three mechanisms -- surface conduction, EOF and EC --- are responsible for the OLC. Experimentally, here the observed OLC, EC and DS are unaffected by modified surface charges, thus ruling out the first two SC mechanisms, in which residual conductivity arises from excess counter-ions screening charged side walls.  In those cases, scalings of the over-limiting conductance with reservoir salt concentration and channel thickness have been predicted \cite{Dydek2011overlimiting} and confirmed experimentally ~\cite{deng2013overlimiting,KimPRL2015}.  It is beyond the scope of this Letter to do the same for EC-driven OLC, but we note that our key model assumption, that $c_d$ is nearly constant during shock propagation, is consistent with previous studies of EC without geometrical confinement \cite{rubinstein2000electroosmotically, zaltzman2007electroosmotic, rubinstein2008direct, Dydek2011overlimiting}.

Our results also hold for negative voltage bias ($\Phi < 0$), where vortices appear at the outer ring, while DS propagates inward at a higher current (SM). It would also be interesting to test predictions of conformal invariance of ion transport in the absence of EC~\cite{Bazantrspa} by studying off-center positions and diverse cross-sectional (elliptical or cloverlike) shapes  \cite{BazantPRE2006} of the wire.
The breakdown of conformal invariance in the transient problem also introduces flexibility to control DS stability~\cite{han2016dendrite}.

Our observations of EC-driven DS are likely to also hold in other configurations, such as the imposed cross flow in shock electrodialysis \cite{deng2013overlimiting, lu2015scalable} and microscale electrodialysis with vortices organized in the depleted region behind a fairly smooth DS \cite{handesalination2013}. This insight may provide guidance to achieve shock electrodialysis in bulk electrolytes without confinement by a charged porous medium, for example in a simple electrodialysis-type stack with only one type of cation membrane, which may enable greater flow rates for continuous and scalable desalination due to the lower hydraulic resistance, albeit with the likely trade-off of lower desalination factor.  Similar phenomena could also be exploited to control electrodeposition.

In conclusion, in a circular channel, a layer of bulk EC vortices appears in the horizontal plane to sustain OLC. The EC-driven depletion layer propagates radially as DS, and the propagation has a remarkable agreement with the proposed model. The EC-driven DS phenomenon may be exploited in new designs of shock electrodialysis for desalination and water purification.


\begin{thebibliography}{22}%
\makeatletter
\providecommand \@ifxundefined [1]{%
 \@ifx{#1\undefined}
}%
\providecommand \@ifnum [1]{%
 \ifnum #1\expandafter \@firstoftwo
 \else \expandafter \@secondoftwo
 \fi
}%
\providecommand \@ifx [1]{%
 \ifx #1\expandafter \@firstoftwo
 \else \expandafter \@secondoftwo
 \fi
}%
\providecommand \natexlab [1]{#1}%
\providecommand \enquote  [1]{``#1''}%
\providecommand \bibnamefont  [1]{#1}%
\providecommand \bibfnamefont [1]{#1}%
\providecommand \citenamefont [1]{#1}%
\providecommand \href@noop [0]{\@secondoftwo}%
\providecommand \href [0]{\begingroup \@sanitize@url \@href}%
\providecommand \@href[1]{\@@startlink{#1}\@@href}%
\providecommand \@@href[1]{\endgroup#1\@@endlink}%
\providecommand \@sanitize@url [0]{\catcode `\\12\catcode `\$12\catcode
  `\&12\catcode `\#12\catcode `\^12\catcode `\_12\catcode `\%12\relax}%
\providecommand \@@startlink[1]{}%
\providecommand \@@endlink[0]{}%
\providecommand \url  [0]{\begingroup\@sanitize@url \@url }%
\providecommand \@url [1]{\endgroup\@href {#1}{\urlprefix }}%
\providecommand \urlprefix  [0]{URL }%
\providecommand \Eprint [0]{\href }%
\providecommand \doibase [0]{http://dx.doi.org/}%
\providecommand \selectlanguage [0]{\@gobble}%
\providecommand \bibinfo  [0]{\@secondoftwo}%
\providecommand \bibfield  [0]{\@secondoftwo}%
\providecommand \translation [1]{[#1]}%
\providecommand \BibitemOpen [0]{}%
\providecommand \bibitemStop [0]{}%
\providecommand \bibitemNoStop [0]{.\EOS\space}%
\providecommand \EOS [0]{\spacefactor3000\relax}%
\providecommand \BibitemShut  [1]{\csname bibitem#1\endcsname}%
\let\auto@bib@innerbib\@empty
\bibitem [{\citenamefont {Probstein}(2003)}]{Probsteinbook}%
  \BibitemOpen
  \bibfield  {author} {\bibinfo {author} {\bibfnamefont {R.~F.}\ \bibnamefont
  {Probstein}},\ }\href@noop {} {\emph {\bibinfo {title} {Physicochemical
  Hydrodynamics}}}\ (\bibinfo  {publisher} {Wiley, New York},\ \bibinfo {year}
  {2003})\BibitemShut {NoStop}%
\bibitem [{\citenamefont {Newman}\ and\ \citenamefont
  {Thomas-Alyea}(2004)}]{Newmanbook}%
  \BibitemOpen
  \bibfield  {author} {\bibinfo {author} {\bibfnamefont {J.}~\bibnamefont
  {Newman}}\ and\ \bibinfo {author} {\bibfnamefont {K.~E.}\ \bibnamefont
  {Thomas-Alyea}},\ }\href@noop {} {\emph {\bibinfo {title} {Electrochemical
  Systems}}}\ (\bibinfo  {publisher} {Wiley, New York},\ \bibinfo {year}
  {2004})\BibitemShut {NoStop}%
\bibitem [{\citenamefont {Schoch}\ \emph {et~al.}(2008)\citenamefont {Schoch},
  \citenamefont {Han},\ and\ \citenamefont {Renaud}}]{schoch2008transport}%
  \BibitemOpen
  \bibfield  {author} {\bibinfo {author} {\bibfnamefont {R.~B.}\ \bibnamefont
  {Schoch}}, \bibinfo {author} {\bibfnamefont {J.}~\bibnamefont {Han}}, \ and\
  \bibinfo {author} {\bibfnamefont {P.}~\bibnamefont {Renaud}},\ }\href@noop {}
  {\bibfield  {journal} {\bibinfo  {journal} {Reviews of Modern Physics}\
  }\textbf {\bibinfo {volume} {80}},\ \bibinfo {pages} {839} (\bibinfo {year}
  {2008})}\BibitemShut {NoStop}%
\bibitem [{\citenamefont {Sonin}\ and\ \citenamefont
  {Probstein}(1968)}]{Sonin1968}%
  \BibitemOpen
  \bibfield  {author} {\bibinfo {author} {\bibfnamefont {A.~A.}\ \bibnamefont
  {Sonin}}\ and\ \bibinfo {author} {\bibfnamefont {R.~F.}\ \bibnamefont
  {Probstein}},\ }\href@noop {} {\bibfield  {journal} {\bibinfo  {journal}
  {Desalination}\ }\textbf {\bibinfo {volume} {5}},\ \bibinfo {pages} {293}
  (\bibinfo {year} {1968})}\BibitemShut {NoStop}%
\bibitem [{\citenamefont {Pu}\ \emph {et~al.}(2004)\citenamefont {Pu},
  \citenamefont {Yun}, \citenamefont {Temkin},\ and\ \citenamefont
  {Liu}}]{pu2004ion-enrichment}%
  \BibitemOpen
  \bibfield  {author} {\bibinfo {author} {\bibfnamefont {Q.}~\bibnamefont
  {Pu}}, \bibinfo {author} {\bibfnamefont {J.}~\bibnamefont {Yun}}, \bibinfo
  {author} {\bibfnamefont {H.}~\bibnamefont {Temkin}}, \ and\ \bibinfo {author}
  {\bibfnamefont {S.}~\bibnamefont {Liu}},\ }\href@noop {} {\bibfield
  {journal} {\bibinfo  {journal} {Nano Letters}\ }\textbf {\bibinfo {volume}
  {4}},\ \bibinfo {pages} {1099} (\bibinfo {year} {2004})}\BibitemShut
  {NoStop}%
\bibitem [{\citenamefont {Kim}\ \emph {et~al.}(2010)\citenamefont {Kim},
  \citenamefont {Song},\ and\ \citenamefont {Han}}]{kim2010nanofluidic}%
  \BibitemOpen
  \bibfield  {author} {\bibinfo {author} {\bibfnamefont {S.~J.}\ \bibnamefont
  {Kim}}, \bibinfo {author} {\bibfnamefont {Y.}~\bibnamefont {Song}}, \ and\
  \bibinfo {author} {\bibfnamefont {J.}~\bibnamefont {Han}},\ }\href@noop {}
  {\bibfield  {journal} {\bibinfo  {journal} {Chemical Society Reviews}\
  }\textbf {\bibinfo {volume} {39}},\ \bibinfo {pages} {912} (\bibinfo {year}
  {2010})}\BibitemShut {NoStop}%
\bibitem [{\citenamefont {Rubin}\ \emph {et~al.}(2016)\citenamefont {Rubin},
  \citenamefont {Suss}, \citenamefont {Biesheuvel},\and\ \citenamefont
  {Bercovici}}]{rubin2016induced-charge}%
  \BibitemOpen
  \bibfield  {author} {\bibinfo {author} {\bibfnamefont {S.}~\bibnamefont
  {Rubin}}, \bibinfo {author} {\bibfnamefont {M.~E.}\ \bibnamefont {Suss}},
  \bibinfo {author} {\bibfnamefont {P.~M.}\ \bibnamefont {Biesheuvel}},\ and\
  \bibinfo {author} {\bibfnamefont {M.}~\bibnamefont {Bercovici}},\ }\href@noop
  {} {\bibfield  {journal} {\bibinfo  {journal} {Physical Review Letters}\
  }\textbf {\bibinfo {volume} {117}},\ \bibinfo {pages} {234502} (\bibinfo {year}
  {2016})}\BibitemShut {NoStop}%
\bibitem [{\citenamefont {Porada}\ \emph {et~al.}(2013)\citenamefont {Porada},
  \citenamefont {Zhao}, \citenamefont {Der~Wal}, \citenamefont {Presser},\ and\
  \citenamefont {Biesheuvel}}]{porada2013review}%
  \BibitemOpen
  \bibfield  {author} {\bibinfo {author} {\bibfnamefont {S.}~\bibnamefont
  {Porada}}, \bibinfo {author} {\bibfnamefont {R.}~\bibnamefont {Zhao}},
  \bibinfo {author} {\bibfnamefont {A.~V.}\ \bibnamefont {Der~Wal}}, \bibinfo
  {author} {\bibfnamefont {V.}~\bibnamefont {Presser}}, \ and\ \bibinfo
  {author} {\bibfnamefont {P.~M.}\ \bibnamefont {Biesheuvel}},\ }\href@noop {}
  {\bibfield  {journal} {\bibinfo  {journal} {Progress in Materials Science}\
  }\textbf {\bibinfo {volume} {58}},\ \bibinfo {pages} {1388} (\bibinfo {year}
  {2013})}\BibitemShut {NoStop}%
\bibitem{nikonenko2014} V. V. Nikonenko, A. V. Kovalenko, M. K. Urtenov, N. D.  Pismenskaya, J. Han, P. Sistat, G. Pourcelly, Desalination \textbf{342}, 85 (2014)
\bibitem{han2014overlimiting} J. H. Han, E. Khoo, P. Bai, and M. Z. Bazant, Sci. Rep. \textbf{4}, 7056 (2014)
\bibitem{han2016dendrite} J. H. Han, M. Wang, P. Bai, F. R. Brushett, and M. Z. Bazant, Sci. Rep. \textbf{6}, 28054 (2016)
\bibitem{andersen2012}  M. B. Andersen, M. van Soestbergen,  A. Mani, H. Bruus, P. M. Biesheuvel and M. Z. Bazant,
Physical Review Letters \textbf{109}, 108301 (2012)
\bibitem [{\citenamefont {Rubinstein}\ and\ \citenamefont
  {Zaltzman}(2000)}]{rubinstein2000electroosmotically}%
  \BibitemOpen
  \bibfield  {author} {\bibinfo {author} {\bibfnamefont {I.}~\bibnamefont
  {Rubinstein}}\ and\ \bibinfo {author} {\bibfnamefont {B.}~\bibnamefont
  {Zaltzman}},\ }\href@noop {} {\bibfield  {journal} {\bibinfo  {journal}
  {Physical Review E}\ }\textbf {\bibinfo {volume} {62}},\ \bibinfo {pages}
  {2238} (\bibinfo {year} {2000})}\BibitemShut {NoStop}%
\bibitem [{\citenamefont {Zaltzman}\ and\ \citenamefont
  {Rubinstein}(2007)}]{zaltzman2007electroosmotic}%
  \BibitemOpen
  \bibfield  {author} {\bibinfo {author} {\bibfnamefont {B.}~\bibnamefont
  {Zaltzman}}\ and\ \bibinfo {author} {\bibfnamefont {I.}~\bibnamefont
  {Rubinstein}},\ }\href@noop {} {\bibfield  {journal} {\bibinfo  {journal}
  {Journal of Fluid Mechanics}\ }\textbf {\bibinfo {volume} {579}},\ \bibinfo
  {pages} {173} (\bibinfo {year} {2007})}\BibitemShut {NoStop}%
\bibitem [{\citenamefont {Rubinstein}\ \emph {et~al.}(2008)\citenamefont
  {Rubinstein}, \citenamefont {Manukyan}, \citenamefont {Staicu}, \citenamefont
  {Rubinstein}, \citenamefont {Zaltzman}, \citenamefont {Lammertink},
  \citenamefont {Mugele},\ and\ \citenamefont
  {Wessling}}]{rubinstein2008direct}%
  \BibitemOpen
  \bibfield  {author} {\bibinfo {author} {\bibfnamefont {S.~M.}\ \bibnamefont
  {Rubinstein}}, \bibinfo {author} {\bibfnamefont {G.}~\bibnamefont
  {Manukyan}}, \bibinfo {author} {\bibfnamefont {A.~D.}\ \bibnamefont
  {Staicu}}, \bibinfo {author} {\bibfnamefont {I.}~\bibnamefont {Rubinstein}},
  \bibinfo {author} {\bibfnamefont {B.}~\bibnamefont {Zaltzman}}, \bibinfo
  {author} {\bibfnamefont {R.~G.~H.}\ \bibnamefont {Lammertink}}, \bibinfo
  {author} {\bibfnamefont {F.~G.}\ \bibnamefont {Mugele}}, \ and\ \bibinfo
  {author} {\bibfnamefont {M.}~\bibnamefont {Wessling}},\ }\href@noop {}
  {\bibfield  {journal} {\bibinfo  {journal} {Physical Review Letters}\
  }\textbf {\bibinfo {volume} {101}},\ \bibinfo {pages} {236101} (\bibinfo
  {year} {2008})}\BibitemShut {NoStop}%
\bibitem{Yossifon2008} G. Yossifon and H. C. Chang, Physical Review Letters \textbf{101}, 254501 (2008).
\bibitem{handesalination2013} R. Kwak, G. F. Guan, W. K. Peng, and J. Han, Desalination \textbf{308}, 138 (2013).
\bibitem{rubinstein2015} I. Rubinstein and B. Zaltzman, Physical Review Letters \textbf{114}, 114502 (2015).
\bibitem [{\citenamefont {Dydek}\ \emph {et~al.}(2011)\citenamefont {Dydek},
  \citenamefont {Zaltzman}, \citenamefont {Rubinstein}, \citenamefont {Deng},
  \citenamefont {Mani},\ and\ \citenamefont {Bazant}}]{Dydek2011overlimiting}%
  \BibitemOpen
  \bibfield  {author} {\bibinfo {author} {\bibfnamefont {E.~V.}\ \bibnamefont
  {Dydek}}, \bibinfo {author} {\bibfnamefont {B.}~\bibnamefont {Zaltzman}},
  \bibinfo {author} {\bibfnamefont {I.}~\bibnamefont {Rubinstein}}, \bibinfo
  {author} {\bibfnamefont {D. S.}~\bibnamefont {Deng}}, \bibinfo {author}
  {\bibfnamefont {A.}~\bibnamefont {Mani}}, \ and\ \bibinfo {author}
  {\bibfnamefont {M.~Z.}\ \bibnamefont {Bazant}},\ }\href@noop {} {\bibfield
  {journal} {\bibinfo  {journal} {Physical Review Letters}\ }\textbf {\bibinfo
  {volume} {107}},\ \bibinfo {pages} {118301} (\bibinfo {year}
  {2011})}\BibitemShut {NoStop}%
\bibitem [{\citenamefont {Deng}\ \emph {et~al.}(2013)\citenamefont {Deng},
  \citenamefont {Dydek}, \citenamefont {Han}, \citenamefont {Schlumpberger},
  \citenamefont {Mani}, \citenamefont {Zaltzman},\ and\ \citenamefont
  {Bazant}}]{deng2013overlimiting}%
  \BibitemOpen
  \bibfield  {author} {\bibinfo {author} {\bibfnamefont {D. S.}~\bibnamefont
  {Deng}}, \bibinfo {author} {\bibfnamefont {E.~V.}\ \bibnamefont {Dydek}},
  \bibinfo {author} {\bibfnamefont {J.}~\bibnamefont {Han}}, \bibinfo {author}
  {\bibfnamefont {S.}~\bibnamefont {Schlumpberger}}, \bibinfo {author}
  {\bibfnamefont {A.}~\bibnamefont {Mani}}, \bibinfo {author} {\bibfnamefont
  {B.}~\bibnamefont {Zaltzman}}, \ and\ \bibinfo {author} {\bibfnamefont
  {M.~Z.}\ \bibnamefont {Bazant}},\ }\href@noop {} {\bibfield  {journal}
  {\bibinfo  {journal} {Langmuir}\ }\textbf {\bibinfo {volume} {29}},\ \bibinfo
  {pages} {16167} (\bibinfo {year} {2013})}\BibitemShut {NoStop}%
\bibitem{Leakymodel2013} E. V. Dydek and M. Z. Bazant, AIChE Journal \textbf{59}, 3539 (2013).
\bibitem{KimPRL2015} S. Nam, I. Cho, J. Heo, G. Lim, M. Z. Bazant, D. J. Moon, G. Y. Sung, and S. J. Kim, Physical Review Letters  \textbf{114}, 114501 (2015).
\bibitem{khoo2018} E. Khoo and M. Z. Bazant, J. Electroanal. Chem. \textbf{811}, 105 (2018). 
\bibitem{ManitheoryLang2009} A. Mani, T. A. Zangle, and J. G. Santiago, Langmuir \textbf{25}, 3898 (2009).
\bibitem{ManiexpLang2009} T. A. Zangle, A. Mani, and J. G. Santiago, Langmuir \textbf{25}, 3909 (2009).
\bibitem [{\citenamefont {Zangle}\ \emph
  {et~al.}(2010{\natexlab{a}})\citenamefont {Zangle}, \citenamefont {Mani},\
  and\ \citenamefont {Santiago}}]{zangle2010theory}%
  \BibitemOpen
  \bibfield  {author} {\bibinfo {author} {\bibfnamefont {T.~A.}\ \bibnamefont
  {Zangle}}, \bibinfo {author} {\bibfnamefont {A.}~\bibnamefont {Mani}}, \ and\
  \bibinfo {author} {\bibfnamefont {J.~G.}\ \bibnamefont {Santiago}},\
  }\href@noop {} {\bibfield  {journal} {\bibinfo  {journal} {Chemical Society
  Reviews}\ }\textbf {\bibinfo {volume} {39}},\ \bibinfo {pages} {1014}
  (\bibinfo {year} {2010}{\natexlab{a}})}\BibitemShut {NoStop}%
\bibitem [{\citenamefont {Mani}\ and\ \citenamefont
  {Bazant}(2011)}]{mani2011deionization}%
  \BibitemOpen
  \bibfield  {author} {\bibinfo {author} {\bibfnamefont {A.}~\bibnamefont
  {Mani}}\ and\ \bibinfo {author} {\bibfnamefont {M.~Z.}\ \bibnamefont
  {Bazant}},\ }\href@noop {} {\bibfield  {journal} {\bibinfo  {journal}
  {Physical Review E}\ }\textbf {\bibinfo {volume} {84}},\ \bibinfo {pages}
  {061504} (\bibinfo {year} {2011})}\BibitemShut {NoStop}%
\bibitem [{\citenamefont {Deng}\ \emph {et~al.}(2015)\citenamefont {Deng},
  \citenamefont {Aouad}, \citenamefont {Braff}, \citenamefont {Schlumpberger},
  \citenamefont {Suss},\ and\ \citenamefont {Bazant}}]{deng2015water}%
  \BibitemOpen
  \bibfield  {author} {\bibinfo {author} {\bibfnamefont {D. S.}~\bibnamefont
  {Deng}}, \bibinfo {author} {\bibfnamefont {W.}~\bibnamefont {Aouad}},
  \bibinfo {author} {\bibfnamefont {W.~A.}\ \bibnamefont {Braff}}, \bibinfo
  {author} {\bibfnamefont {S.}~\bibnamefont {Schlumpberger}}, \bibinfo {author}
  {\bibfnamefont {M.}~\bibnamefont {Suss}}, \ and\ \bibinfo {author}
  {\bibfnamefont {M.~Z.}\ \bibnamefont {Bazant}},\ }\href@noop {} {\bibfield
  {journal} {\bibinfo  {journal} {Desalination}\ }\textbf {\bibinfo {volume}
  {357}},\ \bibinfo {pages} {77} (\bibinfo {year} {2015})}\BibitemShut
  {NoStop}%
\bibitem{lu2015scalable} S. Schlumpberger, N. B. Lu, M. E. Suss, and M. Z. Bazant, Environmental Science and Technology Letters \textbf{2}, 367 (2015).
\bibitem{yaroshchuk2012acis} A. Yaroshchuk, Adv. Coll. Interface Sci. \textbf{183}, 68 (2012).
\bibitem{methodposcharge} PDMS surfaces were positively charged through being immersed into poly(allylamine hydrochloride)
(PAH) solution (1 mg/ml PAH with 0.1 M NaCl) for 2 hours followed after the air plasma treatment.
\bibitem [{\citenamefont {De~Valenca}\ \emph {et~al.}(2015)\citenamefont
  {De~Valenca}, \citenamefont {Wagterveld}, \citenamefont {Lammertink},\ and\
  \citenamefont {Tsai}}]{devalenca2015dynamics}%
  \BibitemOpen
  \bibfield  {author} {\bibinfo {author} {\bibfnamefont {J. C.}~\bibnamefont
  {de~Valenca}}, \bibinfo {author} {\bibfnamefont {R.~M.}\ \bibnamefont
  {Wagterveld}}, \bibinfo {author} {\bibfnamefont {R.~G.~H.}\ \bibnamefont
  {Lammertink}}, \ and\ \bibinfo {author} {\bibfnamefont {P.~A.}\ \bibnamefont
  {Tsai}},\ }\href@noop {} {\bibfield  {journal} {\bibinfo  {journal} {Physical
  Review E}\ }\textbf {\bibinfo {volume} {92}},\ \bibinfo {pages} {031003(R)}
  (\bibinfo {year} {2015})}\BibitemShut {NoStop}%
\bibitem{methodflow} In order to visualize vortex, $0.001 \%$ 1-$\mu$m-diameter fluorescent particles (Invitrogen)
was added into 1 mM aqueous CuSO$_4$ solution, together with $0.1 \%$ Tween80 (Sigma Aldrich) to avoid particle aggregation
\bibitem{AliPRL2016} E. Karatay, M. B. Andersen, M. Wessling, and A. Mani, Physical Review Letters  \textbf{116}, 194501 (2016).
\bibitem{BazantPRE2006} M. Z. Bazant, Physical Review E \textbf{73}, 060601(R) (2006).
\bibitem{Bazantrspa} M. Z. Bazant, Proc. R. Soc. Lond. A \textbf{460}, 1433 (2004).
\end{thebibliography}
\end{document}